\documentclass[10pt,twocolumn]{article}
\usepackage[dvips]{graphicx}
\DeclareGraphicsExtensions{ps,eps}
\usepackage{latexsym}
\def\beq{\begin{equation}}
\def\eeq{\end{equation}}
\def\S{\mbox{S}}
\def\Per{\mbox{Per}}
\def\ACF{\mbox{ACF}}
\def\CCF{\mbox{CCF}}
\def\f{\mbox{F}}
\def\CS{\mbox{CS}}
\def\Coh{\mbox{Coh}}
\begin{document}
\twocolumn[
\begin{center}
{\bf \Large Cross-spectral analysis of physiological\\
                 tremor and muscle activity.\\
             I. Theory and application to unsynchronized EMG\\} \vskip 0.4cm
     {      J. Timmer${}^{1}$,
                M. Lauk${}^{1,2}$,
                W. Pfleger${}^{1}$ and
                G. Deuschl${}^{3}$} \\ \vskip 0.3cm
        {\normalsize ${}^{1}$Zentrum f\"ur Datenanalyse und Modellbildung,
                Eckerstr. 1, 79104 Freiburg, Germany \\              
               ${}^{2}$Neurologische Universit\"atsklinik Freiburg,
               Breisacher Str. 64, 79110 Freiburg, Germany \\
                ${}^{3}$Neurologische Universit\"atsklinik Kiel,
                Niemannsweg 147, 24105 Kiel, Germany} \\ \vskip 0.3cm
        {\small \em (to appear in Biological Cybernetics)}\\ \vskip 0.1cm
\end{center}
\noindent
{\bf We investigate the relationship between the extensor electromyogram (EMG) 
and tremor time series in physiological hand tremor by cross-spectral 
analysis. Special attention is directed to the phase spectrum and the 
effects of observational noise. We calculate the theoretical phase spectrum 
for a second order linear stochastic process
and compare the results to measured tremor data recorded from subjects who
did not show a synchronized EMG activity in the corresponding extensor 
muscle. The results show that physiological tremor is well described by the
proposed model and that the measured EMG represents a Newtonian
force by which the muscle acts on the hand.} \vskip .5cm]

\section {Introduction}
Time series of hand tremor and the related muscle activities of the
flexor and extensor muscles are obtained by measuring the acceleration 
of the hand
(denoted here by ACC) and the surface electromyogram (denoted here by EMG).
The ACC data of physiological tremor have been described as a linear
stochastic process driven by uncorrelated firing motoneurons (Stiles 
and Randall 1967; Randall 1973;
Rietz and Stiles 1974; Elble and Koller 1990; Gantert et al.~1992;
Timmer et al.~1993). The description of physiological tremor by
a linear model is reasonable because linear approximations hold due
to its small amplitude. These linear stochastic processes and their spectral
and cross-spectral
properties were studied exhaustively (Bloomfield 1976; Brockwell and 
Davis 1987; Priestley 1989).
Usually, they are denoted by autoregressive processes, since actual
values are given by a linear combination of past values plus a driving
noise. In terms of physics, these processes are linear
damped oscillators driven by noise.

In the context of linear stochastic processes the relation between two
processes can be analyzed by investigating phase and modulus, i.e.~coherency, 
of the normalized cross-spectrum. Applications of cross-spectral analysis to
EMG and ACC data of physiological tremor are reported in 
(Fox and Randall 1970; Pashda and Stein
1973; Elble and Randall 1976; Stiles 1983; Iaizzo and Pozos 1992). 
Up to now, the coherency and the phase spectrum were investigated only at a 
single frequency. In particular, the phase was always interpreted as a 
time delay between the two processes.

The interpretation of the phase spectrum as a whole is difficult. For
example, as will be shown below, the phase spectrum between EMG
and ACC time series depends only on the mechanical properties of the 
hand and does not allow to draw conclusions on
the dynamics of the driving force, i.e.~the EMG. In general, the phase 
spectrum can only be interpreted under quite strong assumptions about
the interrelation of the processes. 
We discuss those cases relevant for the EMG -- ACC relationship.
Finally, we compare the spectra predicted from the model with those 
estimated from measured data.

The paper is organized as follows: In the next section we briefly
describe the data. In Section \ref{methods} we introduce the
mathematical background for this and a companion paper 
(Timmer et al.~1998). Section \ref{flat_emg} gives theoretical
and empirical results for physiological tremors showing a flat EMG
power spectrum, resulting from unsynchronized muscle activity.
EMG power spectra exhibiting a synchronization and the possible role of 
reflexes are discussed in a companion paper (Timmer et al.~1998).

\section {The data}  
The data were recorded from normal subjects. The recording technique
is described in detail elsewhere (Deuschl et al.~1991). 
Briefly, the time series of the hand tremor (ACC) were measured by a 
light-weight piezo-resistive accelerometer. The sampling rate is 300 Hz. 
The outstretched hand is supported at the wrist. 
We recorded three data sets for each subject, the first with the hand unloaded, 
the second with a 500gr load and the third with a 1000gr load. The weights
were fixed on the belly of the outstretched hand.
External elements as the amplifiers and the piezoresistive sensors produce
additive white observational noise in each recorded time series, uncorrelated
to the measured dynamical process itself.
The variance of the observational noise can be estimated
from the high frequency part of the power spectrum where the contribution of the
tremor oscillation can be neglected. This noise contributes up to 10\% 
of the variance of the recorded data and has a significant effect
on the estimated coherency and phase spectra as will be shown in 
Section \ref {methods}.

The EMG time series (EMG) were measured by surface electrodes 
fixed over the belly of the extensor carpi ulnaris muscle and the flexor 
carpi ulnaris muscle. 
These data represent broad band noise. The information about a
possible synchronization of the muscle activity 
is encoded in a modulation of this noise.
The data were high pass filtered (cut-off frequency 80~Hz) in order
to remove movement artifacts, rectified
in order to obtain time series reflecting the muscle activity 
(Journee 1983) and then low pass filtered (cut-off frequency 150~Hz)
to avoid aliasing. Finally, the signals were digitized and fed into a computer
for off-line analysis.

Like the tremor time series the EMG time series
are contaminated with additive observational noise. Its variance
cannot be estimated analogously to that of the ACC data, since
uncorrelated EMG activity also shows a flat power spectrum at higher
frequencies indistinguishable from that of the observational noise.

\begin{figure}[!h]
\begin{center}
\includegraphics[scale=.53]{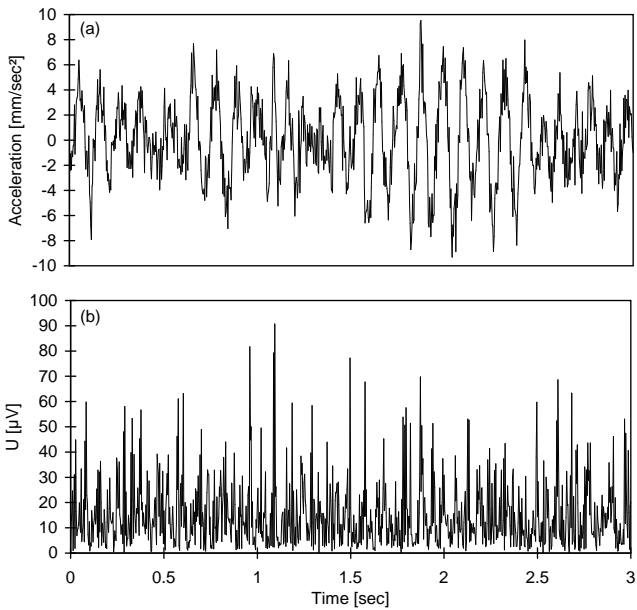}
\end{center}
\caption[]{\label{daten_fig} 
Acceleration of the hand (a) and rectified EMG (b) of
        physiological tremor.}
\end{figure}

Time series from 58 subjects who showed no significant EMG synchronization,
i.e.~a flat spectrum, were examined. The statistical decision of consistency
with a flat spectrum was performed by means of a Kolmogorov-Smirnov test
at the level of confidence p=0.05 (Timmer et al.~1996).
A representative example of such a time series is shown
in Fig.~\ref{daten_fig}.
Time series from 19 subjects with enhanced physiological tremor 
who showed a significant EMG synchronization are analyzed in the
companion paper. In each time series the mean was subtracted and 
all series were scaled to variance one.

\section {Mathematical methods} \label{methods}

In this section we introduce the mathematical methods that will be used
in following sections and the companion paper
 to analyze the simulated and the measured data.
Firstly, we briefly summarize the time and frequency domain properties
of linear stochastic processes before discussing the cross-spectral
estimation and interpretation. Special attention will be payed to
the effects of observational noise which is always present in the
data of physiological tremor and its EMG and renders the interpretation
of the mathematical results more difficult.

\subsection {Linear stochastic processes}

An example of a linear stochastic process is the
autoregressive (AR) process of order $p$ :
\beq 
  x(t) = \sum_{i=1}^{p} a_{i} \, x(t-i) \, + \, \epsilon(t) \quad ,
\eeq
where $ \epsilon(t) $ denotes uncorrelated Gaussian noise with variance
$\sigma^2$. For ease of notation we set the sampling interval to 
unity for the theoretical discussions.
Such a process can be interpreted in terms of physics
 as a combination of relaxators and
damped oscillators (Honerkamp 1993).
For example, an AR process of order 2 with appropriate
parameters  $a_1$ and $a_2$ describes from a physical standpoint 
a linear, damped oscillator with characteristic period $T$ or
 frequency $\omega=1/T$, and relaxation time $\tau$. $T$ and $\tau$ 
are related to the parameters $a_1$ and $a_2$ by:
\begin{eqnarray}
  a_1 & = & 2 \cos \left(\frac{2\pi}{T}\right) \, \exp \,(-1/\tau)\label{a1}\\ 
  a_2 & = & - \exp \,(-2/\tau)\label{a2}  \qquad .
\end{eqnarray}

AR processes can be generalized to the autoregressive moving average
(ARMA) processes by including past driving noise terms in the dynamics.
A more substantial generalization is the linear state space model 
(Honerkamp 1993). It allows one to model explicitly the observational 
noise $\eta(t)$ that covers the dynamical variable $\vec{x}(t)$
which is mapped to the observation by $C$ and 
contributes to the measured $z(t)$:
\begin{eqnarray}
  \vec{x}(t) & = & A \, \vec{x}(t-1) + \vec{\epsilon}(t) \label{zrm1}\\ 
  z(t) & = & C \, \vec{x}(t) + \eta(t) \label{zrm2}  \qquad .
\end{eqnarray}
This model has been applied successfully to physiological tremor
time series (Gantert et al.~1992). If the observational noise is not
modeled explicitly, e.g. by applying an ARMA model, the characteristic
times will be underestimated and statistical tests to decide on
the model order will fail to detect the correct order (K\"onig and 
Timmer 1997). This might explain the high model order reported
earlier (Randall 1973; Miao and Sakamoto 1995).

\subsection {Spectral properties of linear stochastic processes}

The power spectrum $\S(\omega)$ of a mean zero and unit variance process 
$x(t)$ is defined as the Fourier transform 
of the autocorrelation function $\ACF(\tau)$  :
\begin{eqnarray} 
 \ACF (\tau) & = & < x(t)\, x(t-\tau) >  \\
 \S(\omega)  &=  & \frac{1}{2 \pi} \sum_{\tau} \ACF (\tau) 
             e^{-i \omega \tau} \, , \omega \in[-\pi,\pi]
\end{eqnarray}  
with ''$< \, >$'' denoting expectations.
The estimation of the power spectrum is usually based on the Fourier transform
$ \f(\omega) $ and the periodogram  $\Per(\omega)$ of the data :
\begin{eqnarray}
  X(\omega)   & = & \frac{1}{\sqrt{N}} \sum_{t=1}^{N} x(t) \,
     \exp \, (- i\,\omega \, t)                   \\  
  \Per(\omega) & = & |\, X(\omega)\,|^2 
\end{eqnarray} 
and is evaluated at the frequencies:
\beq
  \omega_k = \frac{2\pi k}{N} , \quad  k = -\frac{N}{2}, \ldots,
  \frac{N}{2}-1   \quad . 
\eeq
The expectation of the periodogram is the power spectrum but the
periodogram is not a consistent estimator for the power spectrum
since the standard deviation of this $\chi^2_2$ distributed random variable is
equal to its mean and does not decrease with increasing number of data
(Brockwell and Davis 1987; Priestley 1989):
\beq \label{perverteil}
  \Per (\omega) \sim \frac{1}{2} \S (\omega) \chi^2_2   \quad . 
\eeq
In order to estimate the power spectrum, the periodogram has to be convolved 
by a window function $ W(j) $ of width $2h+1$ :
\beq \label{specschaetz}
   \widehat {\S}(\omega_k) = \frac{1}{2 \pi} \sum_{j=-h}^{h} W(j) \, 
                                \Per(\omega_{k+j}) \quad . 
 \eeq
It is also possible to estimate the power spectrum by averaging
the periodograms of segments of the data or by fitting an AR process
to the data and calculate the spectrum of the fitted process.
General aspects of spectral estimation as well as confidence intervals
are given in (Brockwell and Davis 1987; Priestley 1989). Special aspects
concerning spectral estimation for tremor time series are 
discussed in (Timmer et al.~1996).

For linear processes the power spectrum can be calculated analytically.
In the case of an AR process of order 2 it is given by:
\beq \label{arspec}
 \S(\omega) = \frac{1}{2 \pi} \frac {\sigma^2}
{|1-  a_1 \exp(-i\omega)- a_2 \exp(-2i\omega)|^2} \quad . 
 \eeq
Expressed in terms of $T$ and $\tau$, the power spectrum 
shows for $|\cos (2\pi/T)|\cosh (1/\tau) \leq 1$ a peak at
the frequency:
\beq \label{peakfreq}
    \omega_{peak} = \arccos \left( \cos (2\pi/T) \, \cosh (1/\tau) 
        \right)  \quad . 
\eeq
Therefore, for small $\tau$, the peak of the power spectrum is not located
at the frequency $2\pi/T$. The width of the peak is proportional to $1/\tau$.
If the driving force is characterized by some nontrivial
power spectrum $ \S_{drive}(\omega) $ instead of the constant
spectrum of uncorrelated white noise (\ref{arspec}) changes to:
\beq \label{ardrive} 
  \S(\omega) = \frac{\S_{drive}(\omega)} 
     { | 1 - a_1 \exp (-i \omega)  -  a_2 \exp (-2 i \omega)|^2 } \quad . 
\eeq
\subsection {Cross-spectral analysis}
Analogously to the univariate quantities introduced in the previous
section the cross-spectrum $ \CS (\omega) $ of two zero mean and unit
variance time series $x(t)$ and $y(t)$ is defined as the Fourier 
transform of cross-correlation function $ \CCF (\tau) $  :
\begin{eqnarray} 
  \CCF (\tau) & = & < x(t)\, y(t-\tau) >  \\
  \CS(\omega) & = & \frac{1}{2 \pi} \sum_{\tau} \CCF (\tau) \exp(-i\,\omega \,
  \tau) \\\nonumber
  & = & < X(\omega) \, Y^{\ast}(\omega) > \quad . 
\end{eqnarray} 
Here $^{\ast}$ denotes complex conjugation.
The coherency spectrum $ \Coh(\omega) $ is defined as the modulus of the
normalized cross-spectrum $ \CS (\omega) $ :
\beq
 \Coh(\omega) = \frac{| \CS(\omega) | }{ \sqrt{S_x(\omega) S_y(\omega)} }
           \quad   
\eeq
and the phase spectrum $\Phi(\omega)$ by the representation:
\beq 
\CS(\omega) =  |\CS(\omega)| \, \exp (i\,  \Phi(\omega) )   \quad .  
\eeq
It can be shown that $\Coh(\omega) $ equals one whenever $y(t)$ is a linear
function of $x(t)$. This holds especially for the coherency between an
AR process and its driving noise $\epsilon(t)$ (Brockwell and Davis
1987; Priestley 1989). Thus, the coherency can be interpreted as a
measure of linear predictability. 
The interpretation of the phase spectrum is more difficult. For the following
cases the phase spectrum can be calculated analytically :
\begin{itemize}
\item [-] If the process $y(t)$ is a time delayed version of process
  $x(t)$, i.e.~$y(t)= x(t-\Delta t)$, the phase spectrum is given by a
  straight line with its slope determined by $\Delta t$ :
 \beq \label {delayphase}
  \Phi(\omega) =  \Delta t \, \omega \quad .  
\eeq  
\item [-] If $y(t)$ is the derivative of $x(t)$, i.e. 
        $y(t)=\dot{x}(t)$ a 
        constant phase spectrum of $-\frac{\pi}{2}$ results.
 \beq \label {diffphase}
  \Phi(\omega) = - \pi/2 \quad .  
\eeq  
\item [-] In the case of an AR process of order 2 (AR[2]), the phase spectrum 
  between the driving noise $\epsilon(t)$ and the resulting process is
  given by :
\beq \label {ar2phase}
  \Phi(\omega) = \mbox{arctan} \left( \frac{a_1 \sin {\omega} + 
        a_2 \sin {2 \omega} } { 1- a_1 \cos {\omega} -a_2 \cos {2 \omega} }
        \right)   \quad .  
\eeq
It is important to note that the phase spectrum does not change if 
the driving noise is not a Gaussian white noise process. 
Because of the linearity of the system, eq. (\ref {ar2phase})
holds whenever the driving process shows a broad band power spectrum. 
It might even be chaotic.
If the relaxation time $\tau$ is not smaller than the period $T$,
the phase relation $ \Phi_{discr}(\omega) $ between an (time-discrete)
AR[2] process and its driving noise (\ref{ar2phase}) is in good
approximation related to the well known phase
relation $ \Phi_{cont}(\omega) $ for a
(time-continuous) differential equation of a linear, driven damped oscillator 
by:
\beq \label {ar2cont}
  \Phi_{cont}(\omega) \cong \Phi_{discr}(\omega) + \omega   \quad .  
\eeq

Eq. (\ref{ar2cont}) shows that there is a substantial difference
between a discrete- and continuous-time treatment of the data,
since modeling continuous-time data by discrete time models
yields to an spurious time delay of one unit of the sampling period.
Although, in the case of tremor data, the natural approach would be the
continuous-time version, we choose the discrete-time version because 
the simulation studies become much easier and the mentioned effect is 
easily corrected for. 
\end{itemize}

\begin{figure}[!h]
\begin{center}
\includegraphics[scale=0.4]{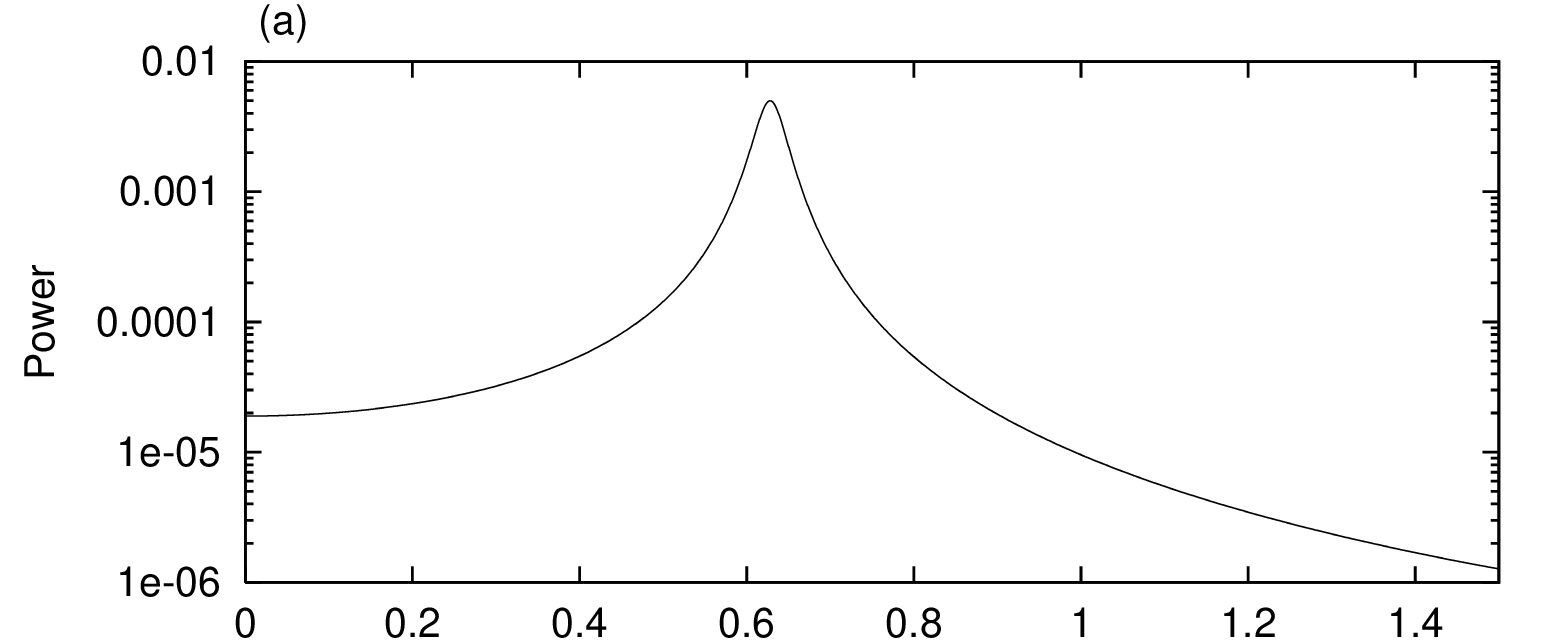}
\includegraphics[scale=0.4]{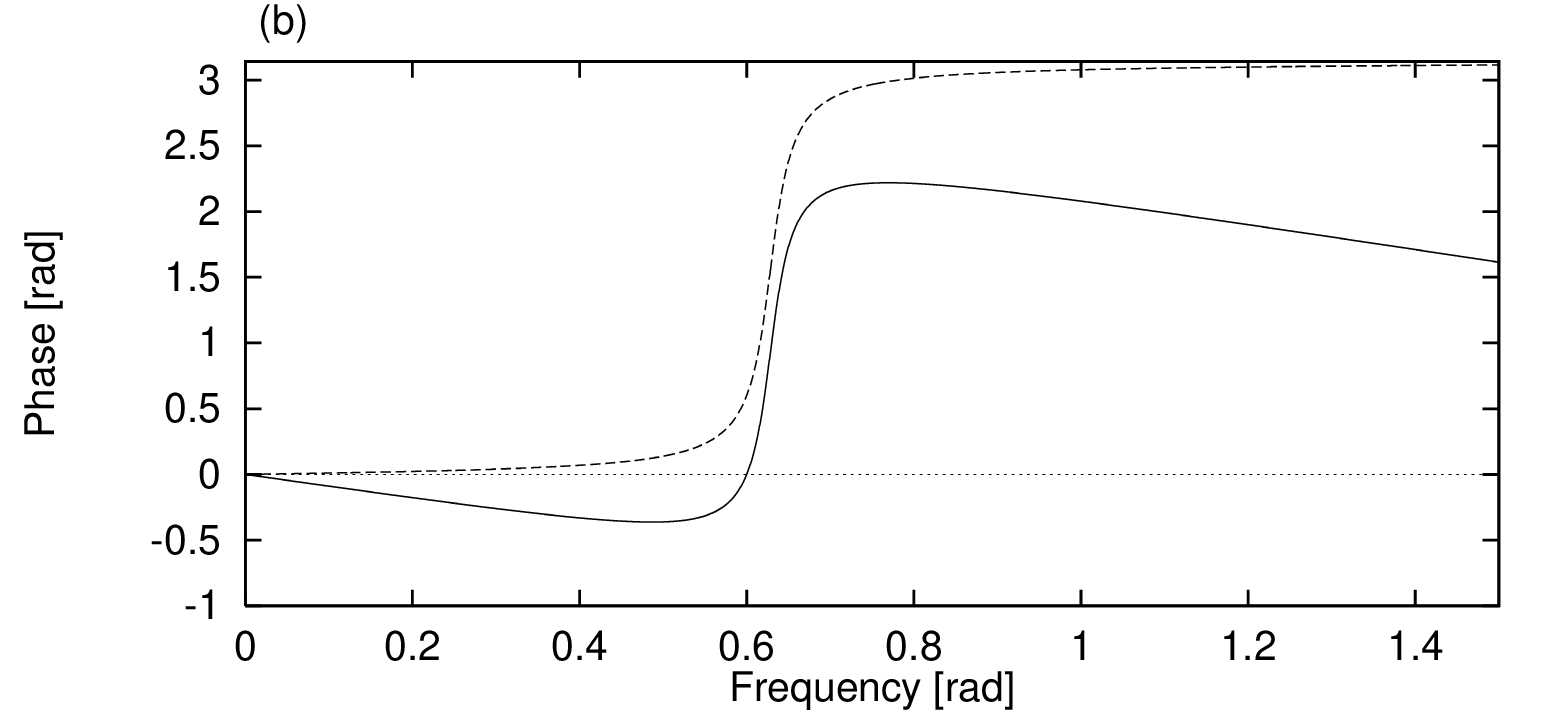}
\end{center}
\caption[]{\label{phasen_fig} 
Power spectrum of an AR[2] process (a). Phase spectrum between
        the AR process and its driving noise (b). The solid line 
        gives the result according to (\ref{ar2phase}), the dashed line the
        result for the continuous time case taking (\ref{ar2cont}) into
        account.}
\end{figure}
Fig.~\ref{phasen_fig}a displays the power spectrum of 
an AR[2] process and Fig.~\ref{phasen_fig}b the phase spectrum between
the AR process and its driving noise. The solid line 
gives the result according to (\ref{ar2phase}), the dashed line shows the
result for the continuous time case taking (\ref{ar2cont}) into account.
Fig.~\ref{phasen_fig} demonstrates that an
interpretation of the phase spectrum $\Phi(\omega)$ for a single
frequency is only possible under strong assumptions about the
relation between the processes. 
In particular, the interpretation of the phase spectrum at a single frequency
$ \omega_0 $, e.g.~the frequency of maximum coherency,
as a time delay by $ \Delta t = \Phi(\omega_0) / \omega_0 $ may be 
erroneous. This situation is similar to the
interpretation of power spectra where in general a certain amount
of power at a certain frequency may not be interpreted as an
oscillator of this variance.
The whole phase spectrum, on the other hand, can provide substantial
informations on the relation between the processes if the empirical 
phase spectrum fits to one of the theoretical phase spectra 
(\ref{delayphase},\ref{diffphase},\ref{ar2phase}) or combinations of them.

The cross-spectrum $\CS(\omega)$ is estimated analogously to 
(\ref{specschaetz}). The critical value $s$ for the null
hypothesis of zero coherency for a significance level
$\alpha$ is given by:
\beq 
  s= \sqrt{1-\alpha^{\frac{2}{\nu-2}}} \quad ,
\eeq
where $\nu$ is determined from the window function $W(j)$ by:
\beq \label {effdof}
  \nu =\frac{2} {\sum_{j=-h}^{h} W^2(j)} \quad .
\eeq
Confidence intervals for the coherency are given in Bloomfield (1976).
Besides the simple case where $y(t)$ and $x(t)$ are indeed 
uncorrelated, at least the following reasons can result in a coherency unequal
one:
\begin{itemize}
\item [-] A nonlinear relationship between $x(t)$ and $y(t)$
\item [-] Additional influences on $y(t)$ apart from $x(t)$
\item [-] Estimation bias due to misalignment (Hannan and Thomson 1971)
\item [-] Observational noise 
\end {itemize}

\begin{figure}[!h]
\begin{center}
\includegraphics[scale=0.52]{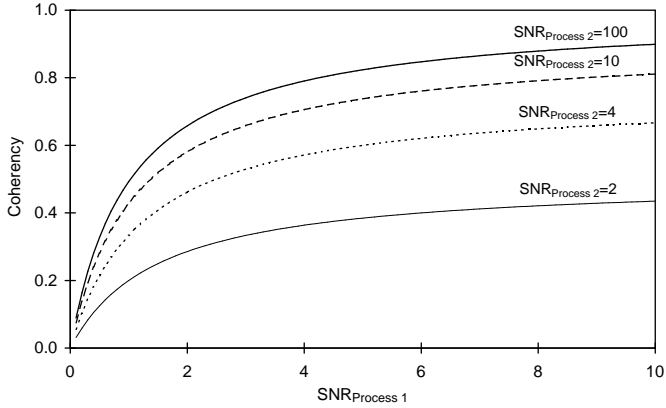}
\end{center}
\caption[]{\label{coherency_noise} 
Coherency between linear related processes in the case
        of observational noise on both processes for different signal 
        to noise ratios (SNR). The abscissa displays the SNR
         for one of the signals. The different curves parameterize the SNR for
        the other process.}
\end{figure}

If $y(t)$ is a linear function of $x(t)$ but the
measurement of $y(t)$ is covered by white observational noise of
variance $\sigma_{ob}^2$ the coherency is given by (Brockwell and Davis
1987) :
\beq 
\Coh(\omega)= \sqrt{\frac{S_y(\omega)}{S_y(\omega) +
    \sigma_{ob}^2}} \quad .  
\eeq
Thus, the coherency is a function of the frequency dependent signal
to noise ratio.
For the general case of observational noise on both processes the coherency
is given by:
\beq \label {coh_noise}
\Coh (\omega)= \sqrt{ 1 - \frac{\sigma_x^2 S_y + S_x \sigma_y^2 +
    \sigma_x^2 \sigma_y^2} {(S_x + \sigma_x^2)(S_y+\sigma_y^2)} } \quad ,
\eeq
where the argument $\omega$  was suppressed on the right hand for ease 
of notation and $\sigma_x^2$ and $\sigma_y^2$ denote the constant 
power spectra of the observational noise. 
Fig.\ref {coherency_noise} illustrates (\ref{coh_noise}) for
\marginpar{please locate Fig.\ref{coherency_noise} near here}
different signal to noise ratios of both processes. 
This might partially explain the findings
of Stiles (1980) and Lenz et al.~(1988), who report a correlation
between peak power and coherency at the peak frequency
as an effect of observational noise. If we assume a constant amount of 
observational noise the peak power is correlated with the signal to noise 
ratio. 

Eq. (\ref{coh_noise}) is of particular interest since the variance of the
estimator for the phase spectrum $\widehat{\Phi}(\omega)$ depends on the
coherency (Priestley 1989):
\beq 
\label{phasenfehler} 
var(\widehat{\Phi}(\omega)) = \frac{1}{\nu}
\left( \frac{1}{{Coh}^2(\omega)} -1 \right) \quad , 
\eeq
where $\nu$ is given by (\ref{effdof}). Eq.~\ref{phasenfehler} holds
if the coherency is significantly larger than zero. For a  
coherency towards zero, the distribution of the estimated phase approaches the
uniform distribution in $[-\pi, \pi]$. Therefore, 
the phase spectrum cannot be estimated reliably in the case of small
coherency. 

\begin{figure}[!h]
\begin{center}
\includegraphics[scale=0.46]{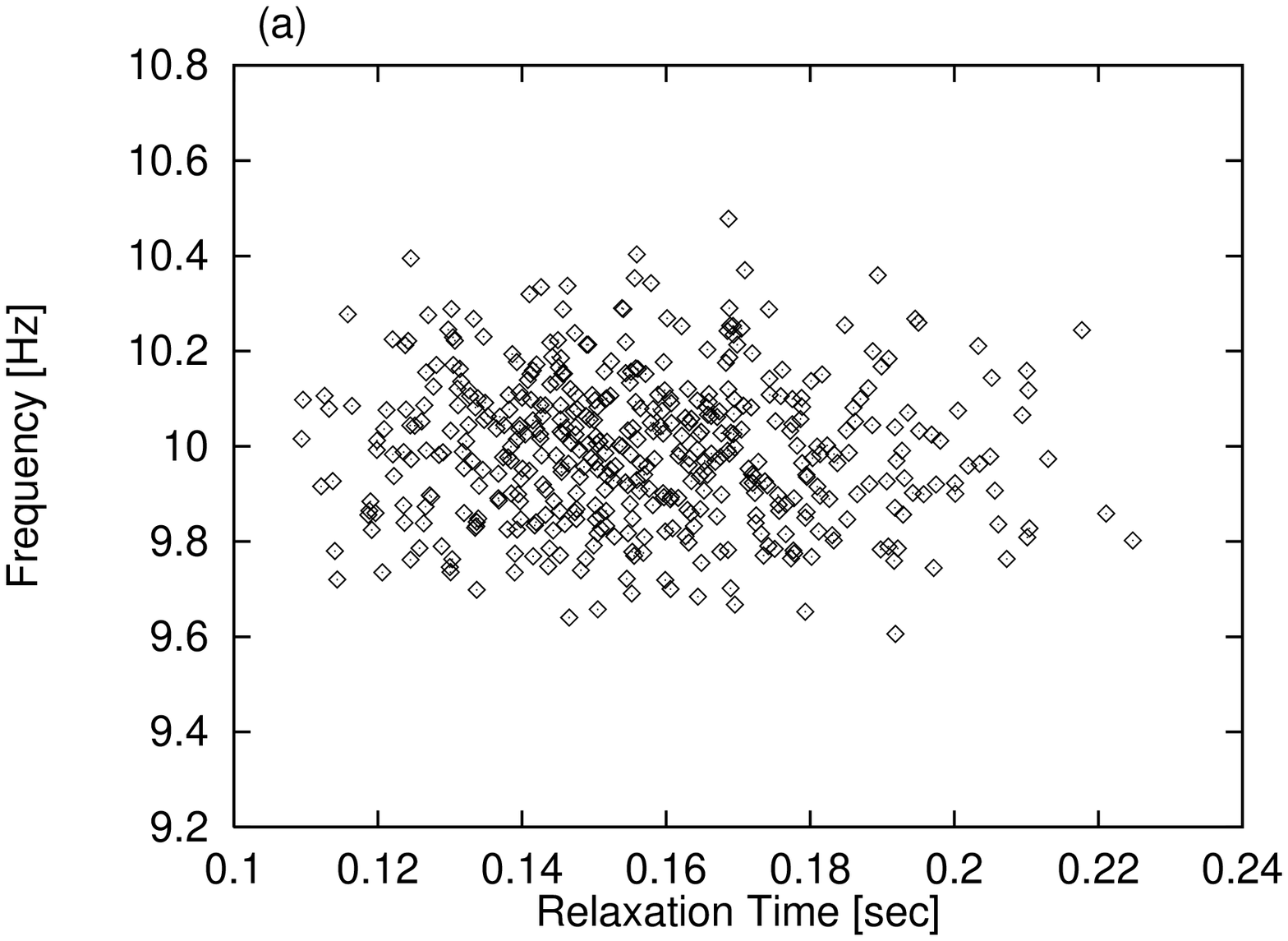}
\includegraphics[scale=.46]{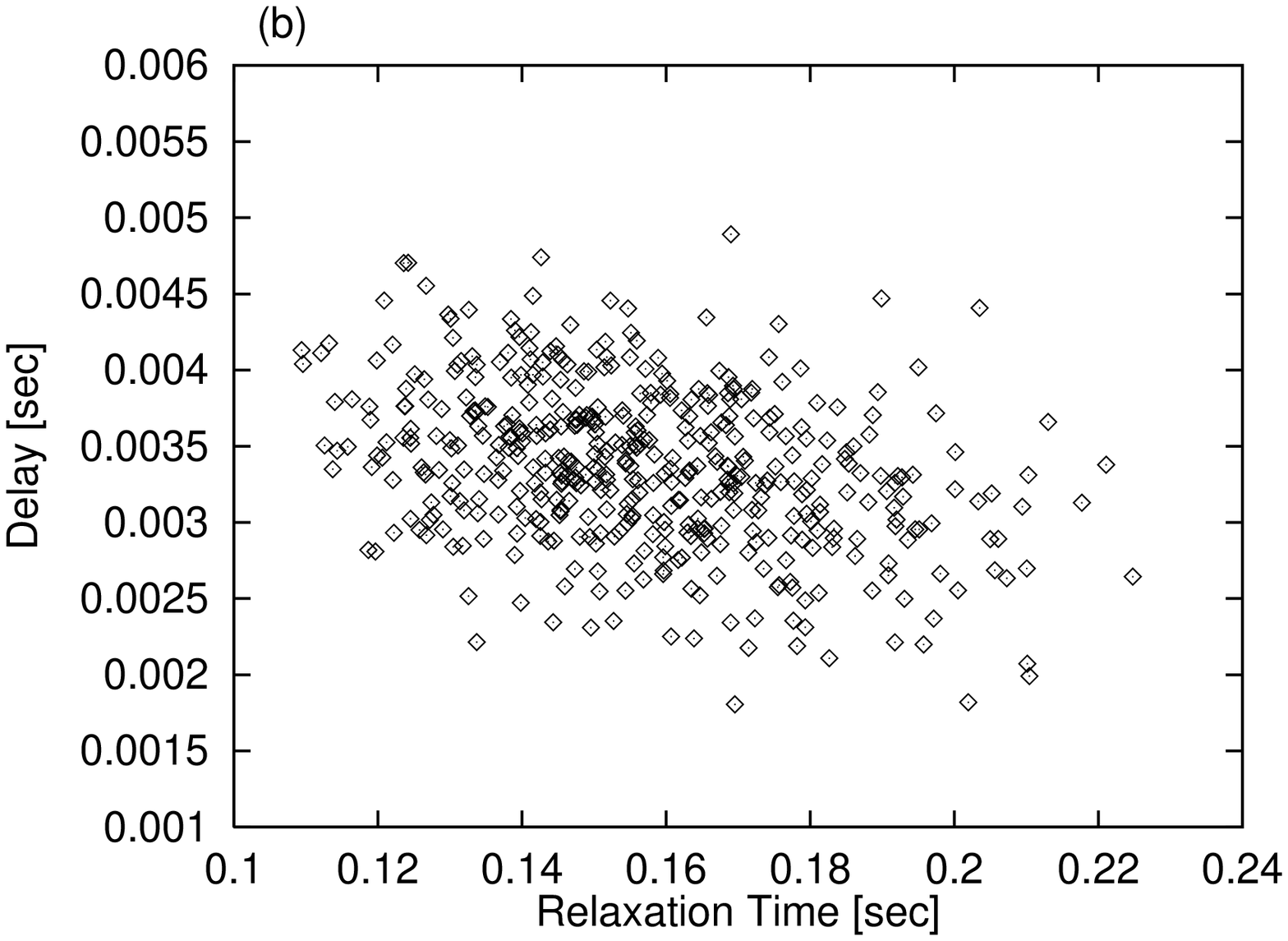}
\end{center}
\caption[]{\label{MC_results} 
Scatter plots of estimated frequency and relaxation
        time $\tau$ (a) and delay $\Delta t $ and relaxation time
        $\tau$ (b) from the Monte Carlo study.}
\end{figure}

Using (\ref {phasenfehler}), theoretical phase spectra 
like (\ref{delayphase}) or (\ref {ar2phase}) can be fitted 
to estimated phase spectra by a maximum likelihood procedure.
We used the Levenberg--Marquardt algorithm (Press et al.~1992).
This algorithm provides confidence limits for the estimated 
parameters, that are asymptotically valid. The asymptotic results
hold in the finite case when the parameter estimates are Gaussian. 
In order to test
whether this condition applies in our case we performed a Monte Carlo
simulation for an AR[2] process under conditions analogous to those 
observed in the empirical data. The variance of the driving noise,
i.e.~the unsynchronized EMG, was set to unity, the frequency
of the AR process was 10 Hz, the relaxation time 0.1 s, i.e.~one 
period. For a sampling frequency of 300 Hz, according to
 (\ref{a1},\ref{a2}) the parameters of the
AR[2] process are $a_1=1.8922$ and $a_2= -0.9355$. Gaussian
observational noise was added to both processes to obtain a signal to
noise ratio of 10:1.
Fig.~\ref {MC_results} shows scatter plots the estimated frequency, 
relaxation time $\tau$ and delay $\Delta t $  for 500 realizations 
of the process.
The Gaussianity of the estimates is clearly visible. 
Furthermore the estimates are uncorrelated.  This is plausible since
the period $T$ determines the frequency at which the phase spectrum is
vastly varying, whereas the
relaxation time $\tau$ is related to the steepness of the phase spectrum
at the that frequency. The estimated delay time
corresponds to the sampling period of $0.0033$ s due to 
(\ref{ar2cont}). The variances and the covariance of zero
are consistent with the results from the Levenberg--Marquardt
algorithm. Note that the relative error in $\tau$ is
much larger than that in $T$. The goodness-of-fit is judged by
the $\chi^2$ statistic (Press et al.~1992).

The peak frequencies of the power spectra were estimated by the 
frequency of maximum power. A bootstrap method to obtain confidence regions 
for the true peak frequency
will be described elsewhere (Timmer et al.~1997). Briefly, many
periodograms are simulated from the estimated power spectra by the
relation (\ref{perverteil})
and the power spectrum is re-estimated. The quantiles of distribution of 
the peak frequencies from the re-estimated power spectra yield a confidence
region for the estimated peak frequency.

If two linear processes with autocorrelation functions  $ \ACF_1(t) $
and $ \ACF_2(t) $ are independent, the estimated cross-correlation 
function $  \widehat {\CCF} (\tau)$ is Gaussian distributed as :
\beq \label {ccfconfi}
  \widehat {\CCF} (\tau) \sim {\cal N} \left(0, N^{-1} \sum_t \ACF_1(t)
  \, \ACF_2(t)\right) \quad .
\eeq
Therefore, independence is difficult to infer from the cross-correlation
function since the confidence interval depends on the autocorrelation
functions of both processes that are in general not known. Only if
one of the processes is white noise, the $ 95 \% $ confidence interval for 
a zero cross-correlation function is given by $ \pm 1.96 N^{-1} $. 
Furthermore, the estimated cross-correlation function is not
uncorrelated for different lags. The covariance is given by :
\beq \label {ccfcov}
  \left( \widehat {\CCF} (\tau_1) \widehat {\CCF} (\tau_2) \right) =
   \frac{1}{N} \sum_t \ACF_1(t) \ACF_2(t+\tau_2-\tau_1) .
\eeq
Again, the autocorrelation functions of both processes enter the equation
 (Brockwell and Davis 1987).
If, for example, one process is white noise and the other is
oscillating the cross-correlation function will show an oscillating
behavior, even if the processes are independent. If the processes 
are not independent, (\ref{ccfconfi},\ref{ccfcov}) contain also the true 
cross-correlation function (Bartlett 1978), rendering the interpretation 
even more difficult.

In the following section we apply the methods introduced above to
measured data of physiological tremor without 
synchronization in the EMG. Whether and how cross-spectral analysis 
can contribute to decide about
the origin of synchronized EMG in the case of (enhanced) physiological tremor
is discussed in a companion paper (Timmer et al.~1998).
In both cases the flexor EMG appeared to have a negligible contribution to
the ACC data, i.e.~the coherency spectrum is most often consistent with 
the hypothesis that the processes are uncorrelated. Whenever there was
a significant coherency it was invoked by cross talk from the 
extensor EMG. The amount of cross talk can be estimated from the
discontinuity at lag zero of the cross-correlation function because of
its instantaneous effect. The 
dominant contribution of the extensor is plausible since it is the
anti gravity muscle. Thus, only the extensor EMG is considered
in the analysis.

\section {Results} \label{flat_emg}
\begin{figure}
\begin{center}
\includegraphics[scale=0.38]{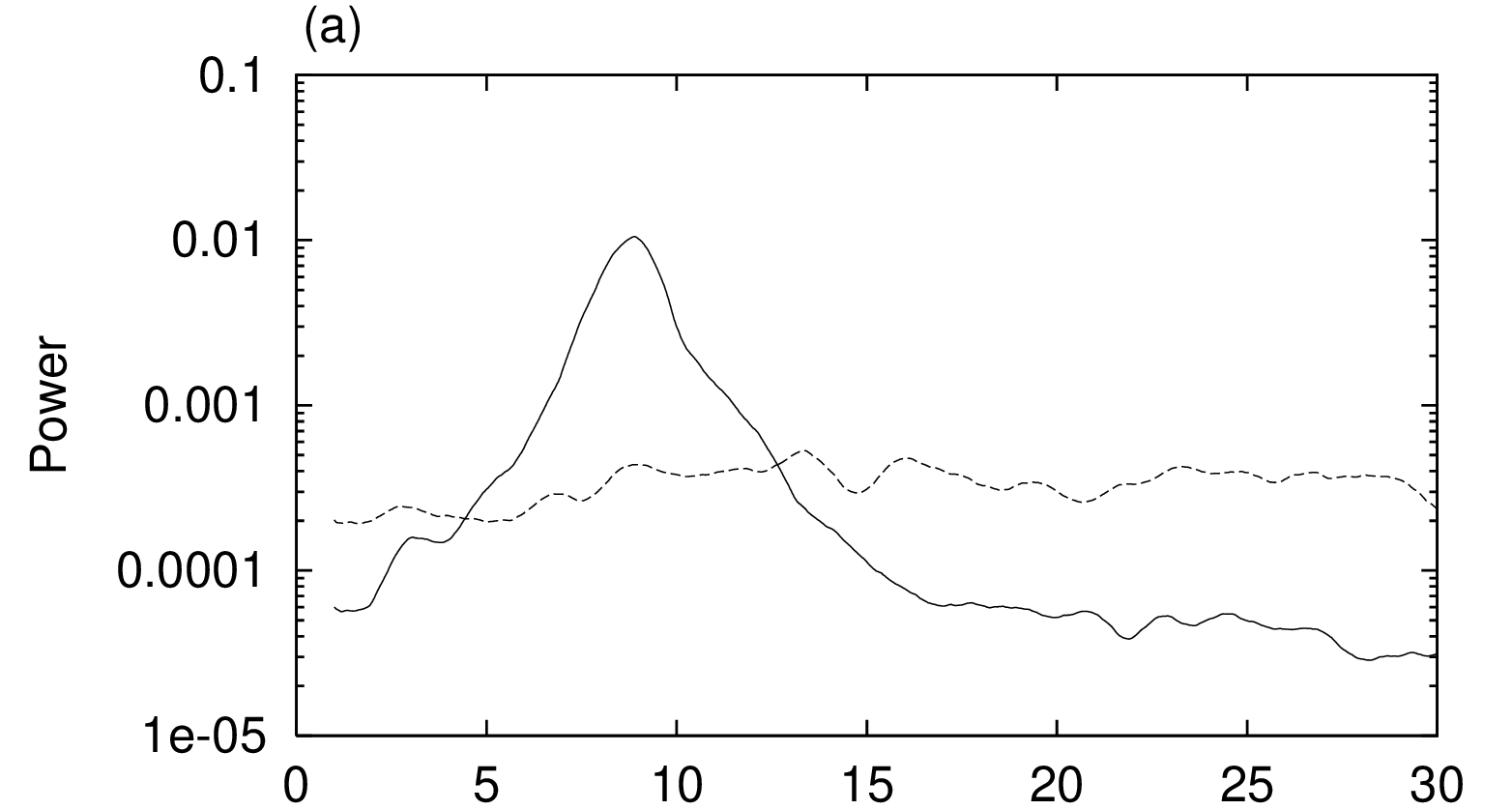}
\includegraphics[scale=.38]{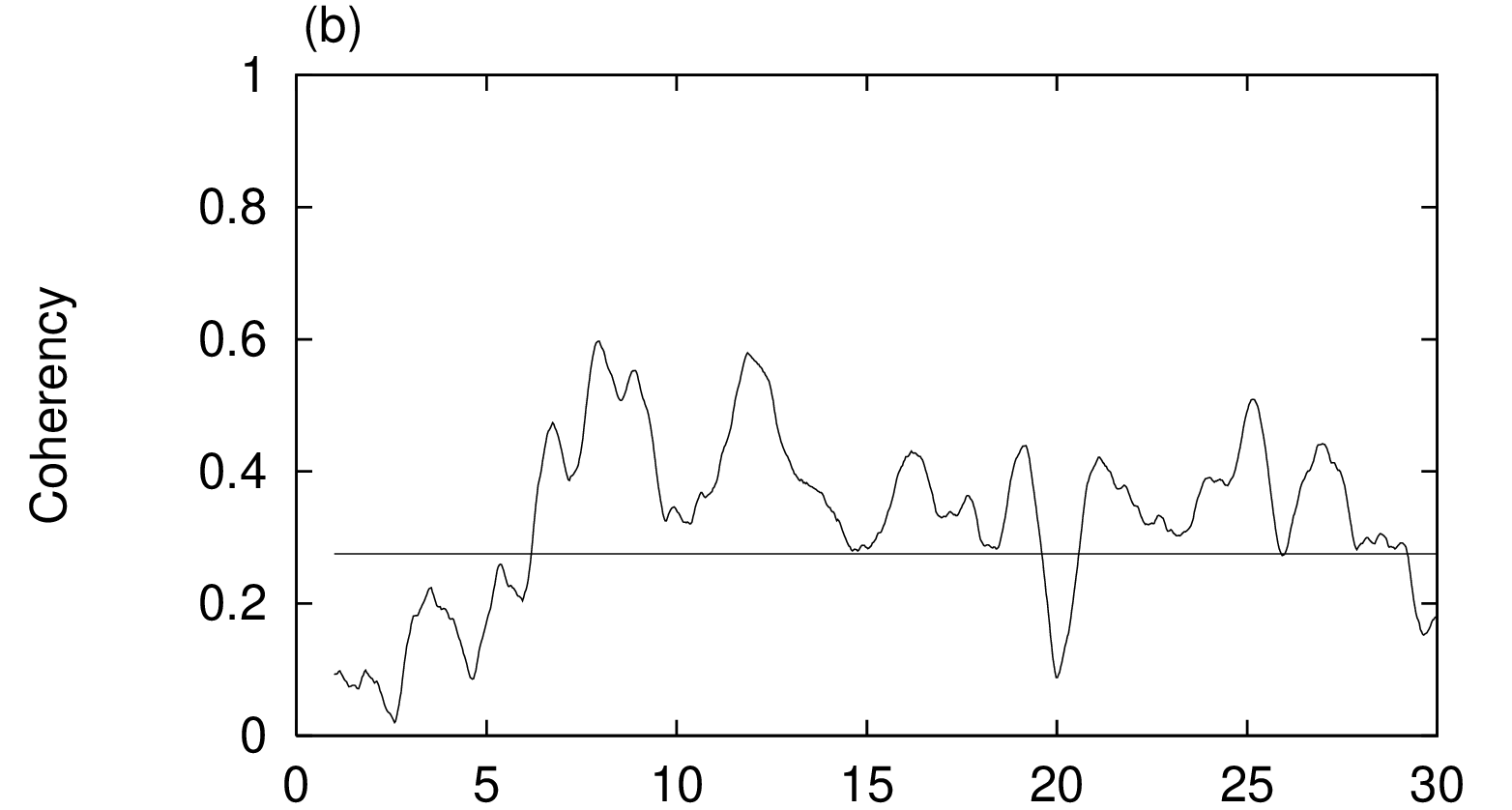}
\includegraphics[scale=.38]{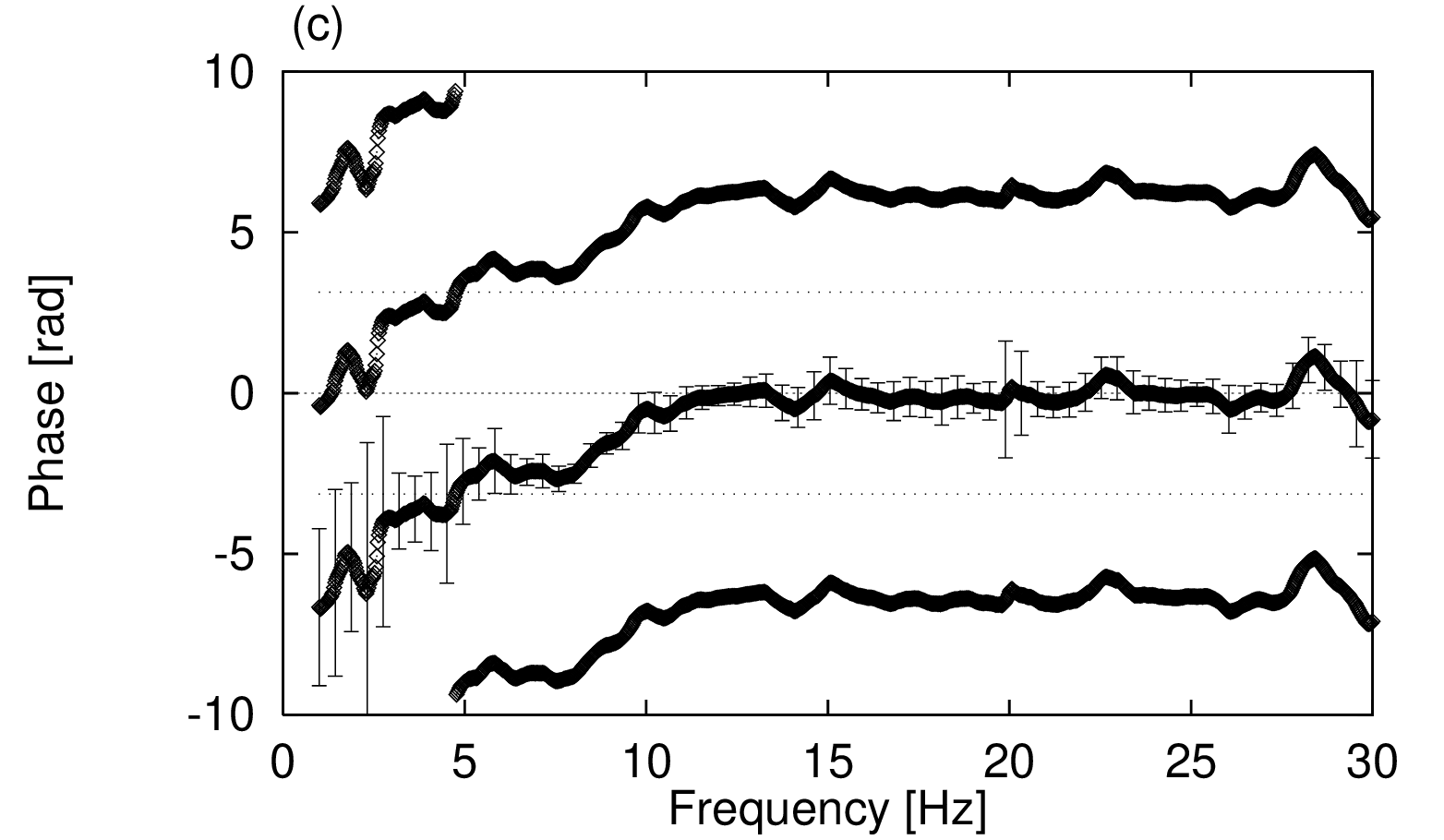}
\includegraphics[scale=.38]{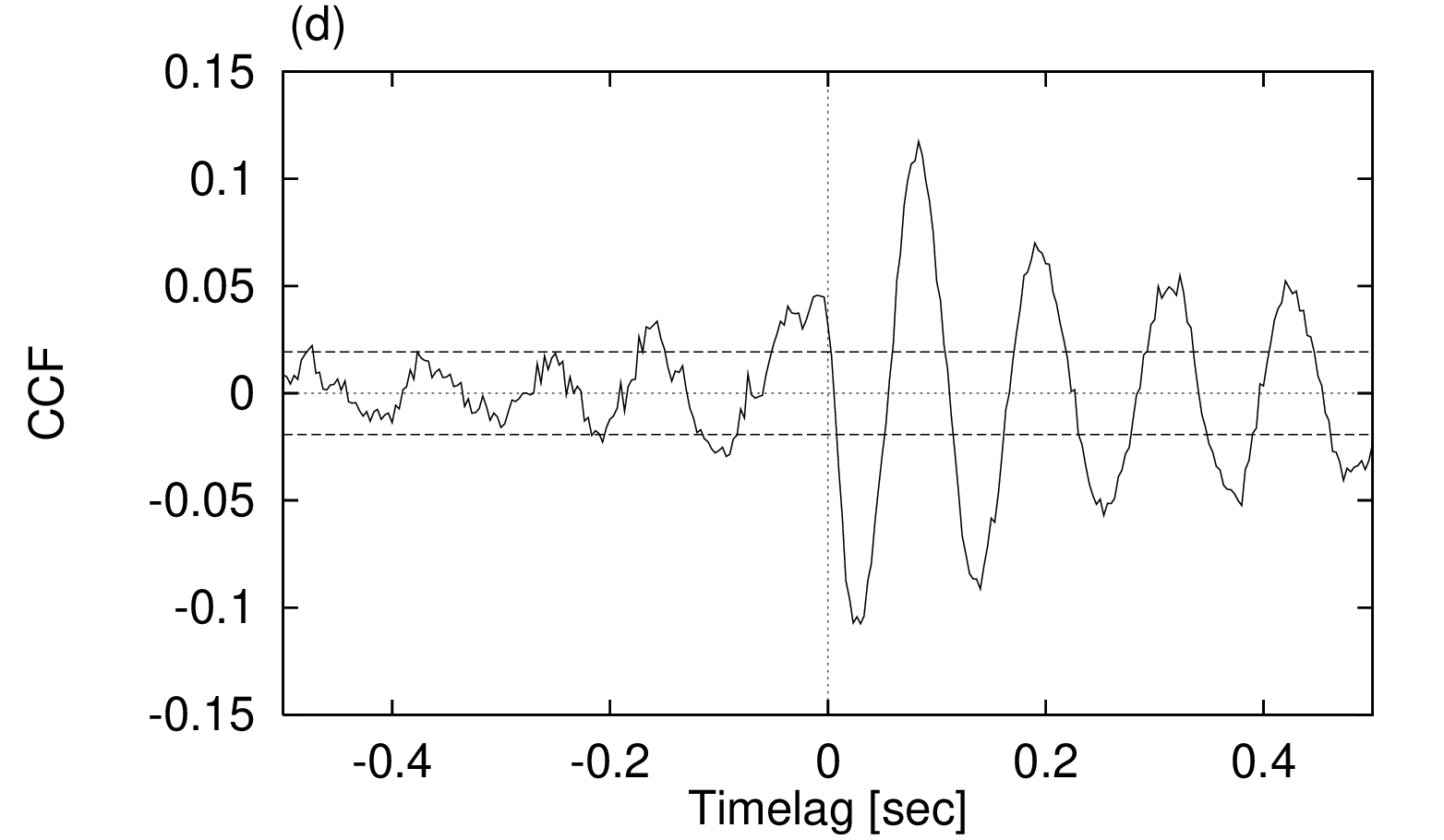}
\end{center}
\caption[]{\label{flat_emg_fig} 
Results for a physiological tremor without EMG synchronization.
      (a): power spectra (EMG: dashed line, ACC: solid line),
      (b): coherency spectrum, the straight line 
       represents the 5\% significance level for the hypothesis of zero 
      coherency, (c): phase spectrum with 95\% confidence intervals, 
        (d): cross-correlation function,
        the dashed lines display the 5 \% significance level for zero
        cross-correlation assuming that at least one of the processes 
        in consistent with white noise. Confidence intervals for the
     power spectra and the coherency are not displayed for reasons of 
        clarity.}
\end{figure}

It was frequently observed that a tremor appears even without synchronization
in the EMG. This was interpreted as a resonance
phenomenon and described by an AR process (Stiles and Randall 1967, 
Randall 1973, Gantert et al.~1992). Fig.~\ref{flat_emg_fig} shows the results
of the spectral and cross-spectral analysis for the  data displayed in
Fig.~\ref{daten_fig}. Fig.~\ref{flat_emg_fig}a shows the 
corresponding spectra, Fig.~\ref{flat_emg_fig}b the coherency spectrum, 
Fig.~\ref{flat_emg_fig}c the  phase spectrum and
Fig.~\ref{flat_emg_fig}d the cross-correlation function estimated as
 described in Section 3.  The straight line in Fig.~\ref{flat_emg_fig}b 
 represents the 5\% significance level for the hypothesis of zero
coherency. The dashed line in Fig.~\ref{flat_emg_fig}d gives the
5\% significance level for the hypothesis of zero cross-correlation
assuming that at least one of the processes is white noise according
to (\ref{ccfconfi}). 
The phase spectrum is shown $2 \pi $ periodically for a range of $\pm 3 \pi$. 
The confidence regions of $2\sigma$ are only given for the central curve. 

The coherency spectrum seems to exhibits two peaks at approximately
8 and 12 Hz. Taking the confidence regions for the true coherency
into account which are not shown for sake of clarity reveals that
these peaks are not significant, but represent a single peak
in the region 7 to 14 Hz. The fact that the coherency spectrum shows its 
maximum values in the region of the peak 
of the ACC power spectrum, can be explained by (\ref{coh_noise})
since EMG and ACC data are contaminated with noise.
Compared to Fig.~\ref{phasen_fig} the phase spectrum of the data is
shifted by $\pi$.
Since we measured the acceleration instead of the position this results
from applying (\ref{diffphase}) twice to the ACC data. For frequencies
below 3 Hz the small coherency and, therefore, the large errors of the 
estimated phase disable its interpretation. Note that the cross-correlation 
function shows a periodic
structure also for negative time lags. Although they are 
statistically not significant, one could speculate whether they give
evidence for some kind of reflex feedback.

\begin{figure}
\begin{center}
\includegraphics[scale=0.38]{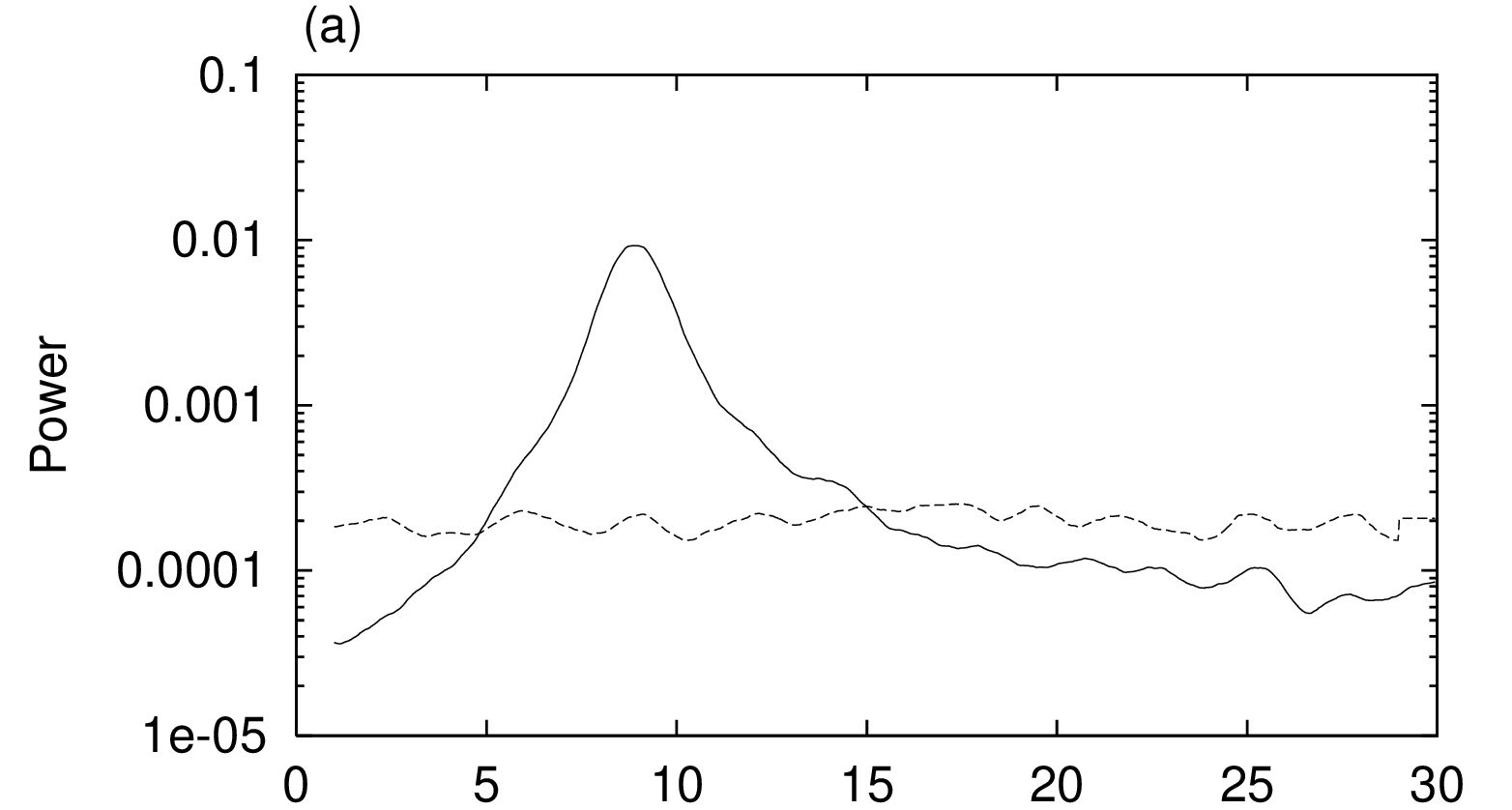}
\includegraphics[scale=.38]{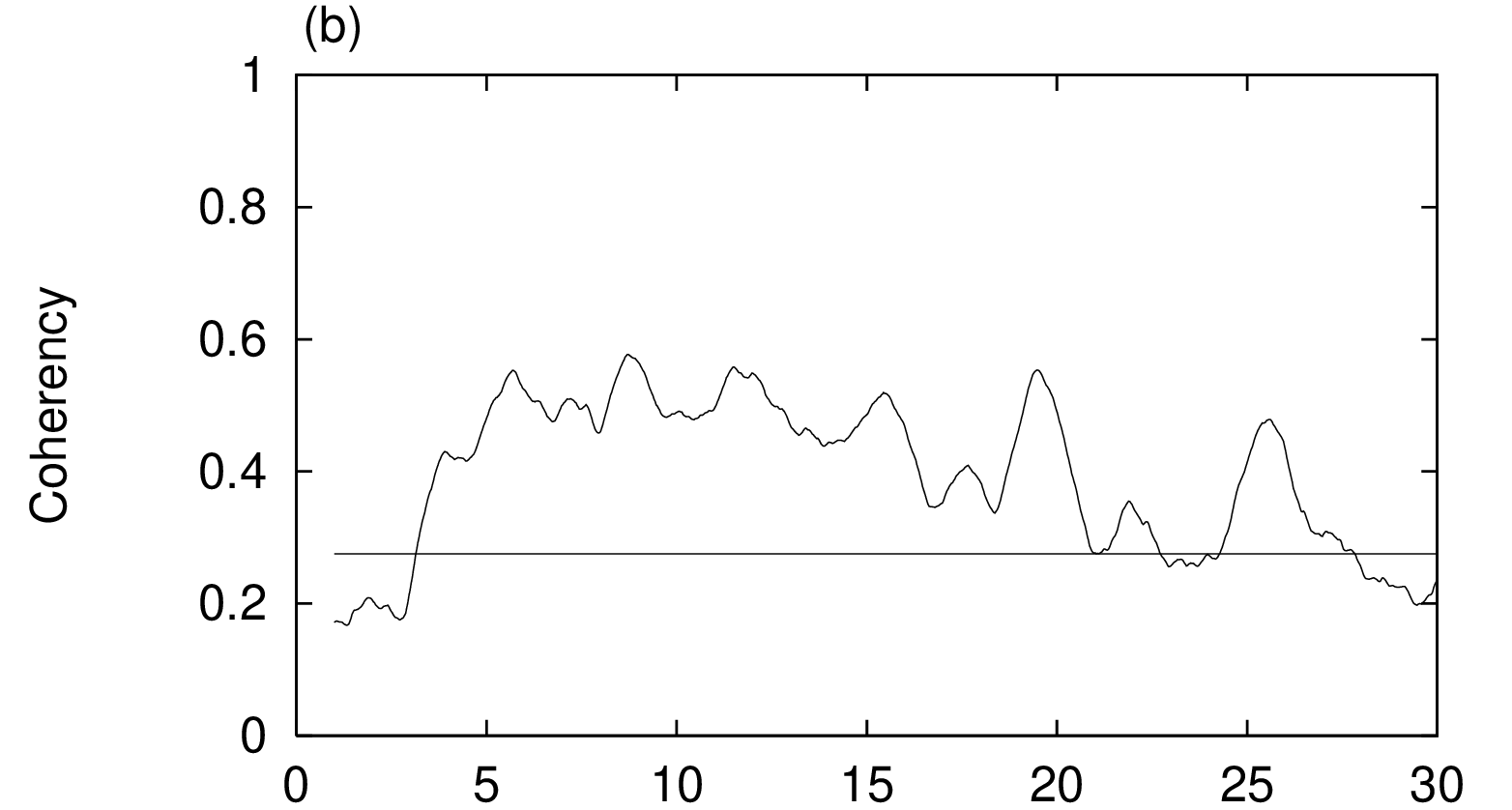}
\includegraphics[scale=.38]{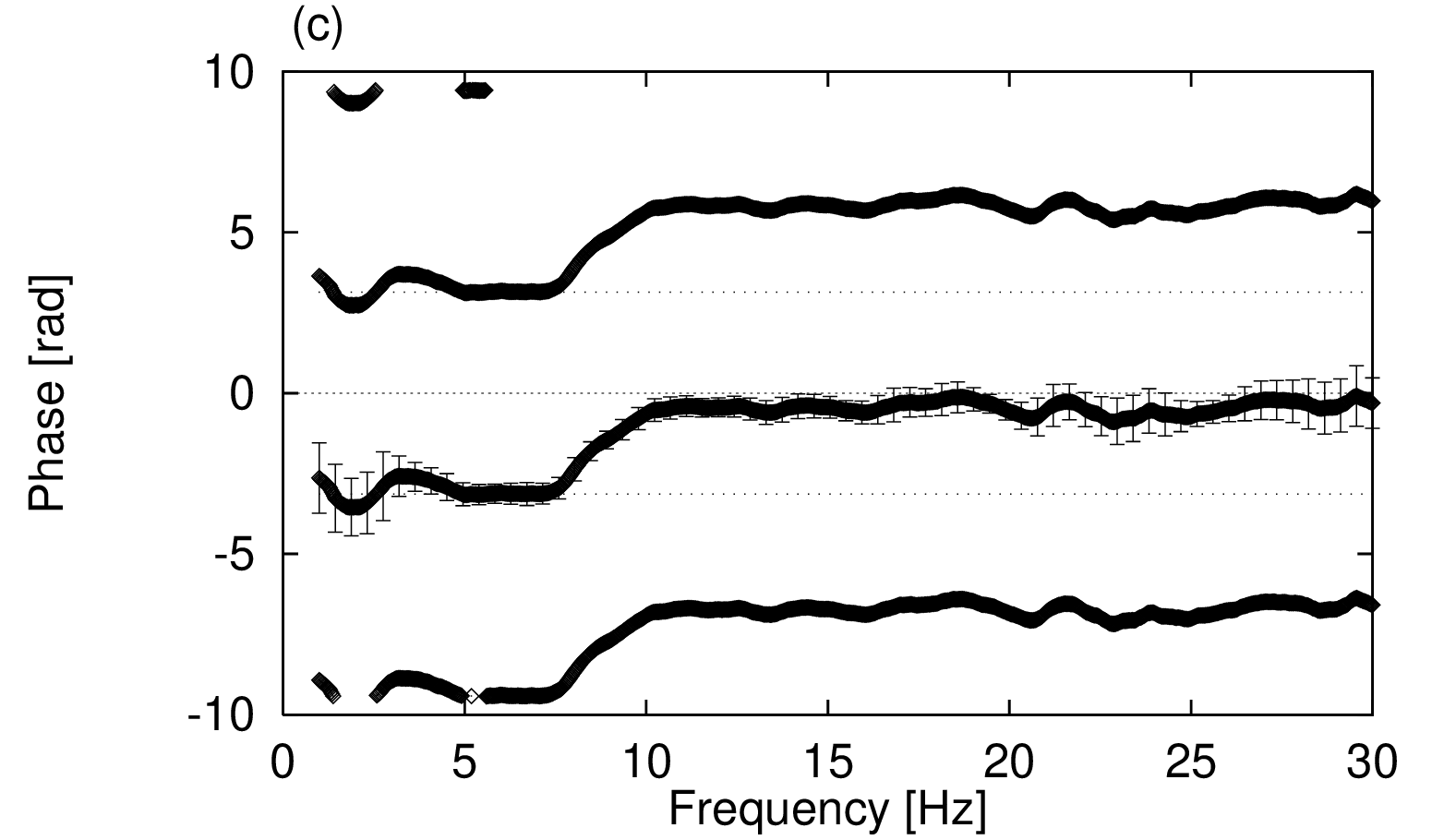}
\includegraphics[scale=.38]{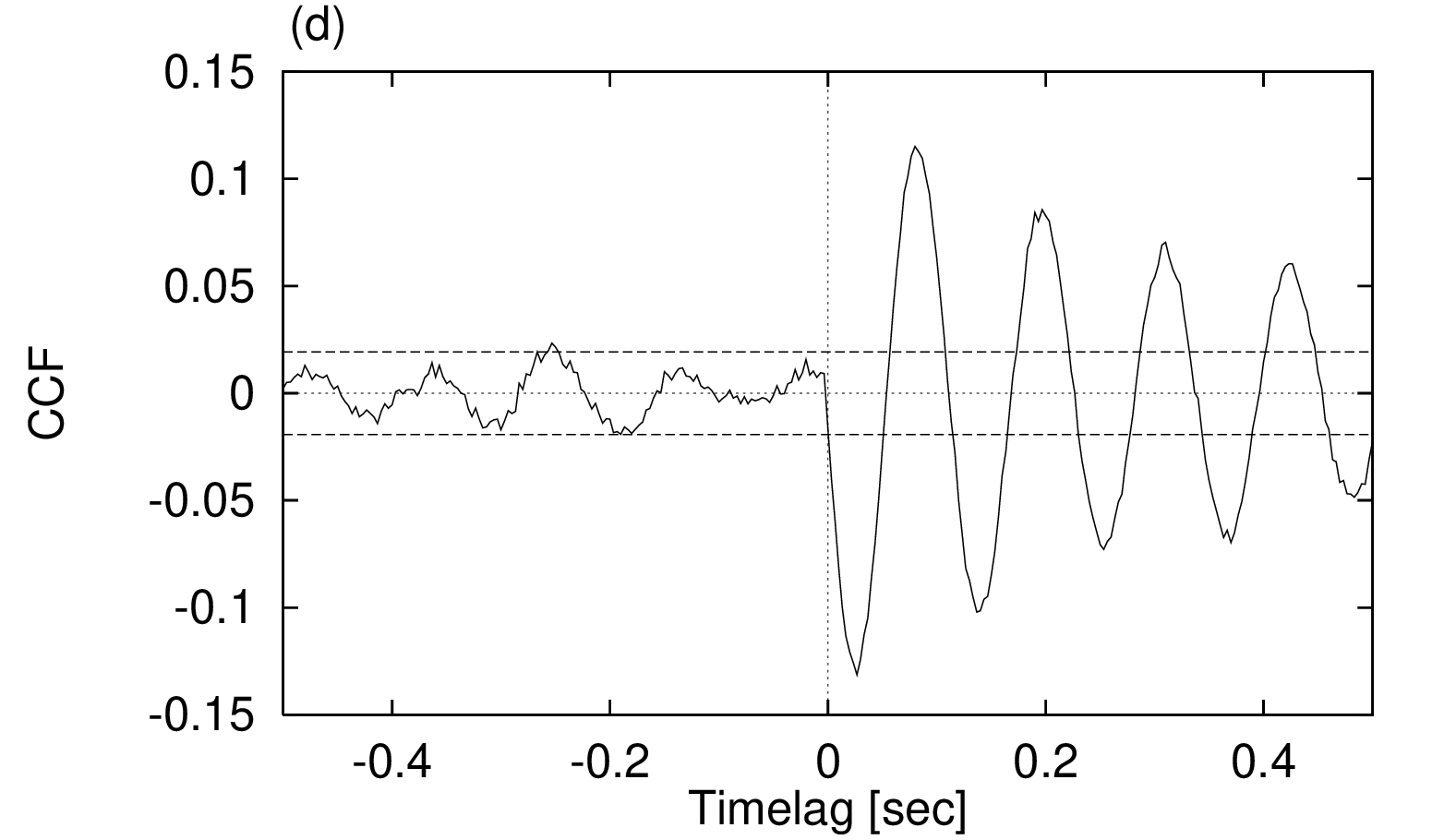}
\end{center}
\caption[]{\label{flat_emg_fit} 
Results for a linear model fitted to the data of
      Fig.~\ref{flat_emg_fig}.}
\end{figure}

In the frame of linear stochastic processes (without reflex feedback)
the different spectra and the cross-correlation function
should be determined by the following six parameters. 
\begin{itemize}
\item The characteristic times $T$ and $\tau$, determining the
        parameters  $ a_1 $ and $ a_2 $ of the AR
        process modeling the mechanic properties of the musculosceletal
        system.
\item A possible time delay $\Delta t$ between the EMG an ACC data.
\item The variance $var_{EMG}$ of the white noise $\epsilon (t) $ modeling
        the asynchronous EMG activity.
\item The variances $var_{obs.\,ACC}$ and $var_{obs.\,EMG}$ of the
        observational noises $\eta_i(t)$.
\end{itemize}
Denoting the EMG by $y(t)$, the movement of the hand by $x(t)$ and the 
measured values by the subscript $m$ the model reads :
\begin{eqnarray}
y(t) & = & \epsilon (t)  
        , \qquad \epsilon (t) \sim {\cal N} (0,var_{EMG}) \\
x(t) & = & a_1 \, x(t-1)  + a_2 \, x(t-2) + y(t-\Delta t ) \\
y_m(t) & =&y(t) + \eta_1(t) \\\nonumber
          & &     \qquad \qquad \eta_1(t) \sim {\cal N} (0,var_{obs.\,EMG})\\
x_m(t) & =&x(t) + \eta_2(t) \\\nonumber
          & &      \qquad \qquad \eta_2(t) \sim {\cal N} (0,var_{obs.\,ACC})
\end{eqnarray}
By
(\ref{a1},\ref{a2},\ref{zrm1},\ref{zrm2},\ref{arspec},\ref{delayphase},\ref{ar2phase},\ref{coh_noise})
we fitted the parameter to the data. First, we fitted the phase spectrum
without taking a possible time delay into account. This resulted in 
an inappropriate fit. Only the inclusion of a time delay according
to (\ref{ar2cont}) into the model gave a fit consistent with the data. 
Note, that this time delay of one sample unit does not reflect 
a time delay between the processes under investigation.
It results from using a time-discrete model to describe an originally 
time-continuous process as discussed in Section 3.3.
A realization of the fitted model and the estimated spectra are displayed in
Fig.~\ref{flat_emg_fit}. Taking into account the errors of all
estimated quantities, it  shows good quantitative agreement
with the empirical results of Fig.~\ref{flat_emg_fig}.
The decreasing coherency for the high frequencies 
due to the frequency dependent signal to noise ratio and the
resulting errors in the phase spectrum are well reproduced.
For the low frequencies the coherency of the model seems be larger
than that of the data. This phenomenon was confirmed in many
data sets. The discrepancy between the data simulated by the model
and the measured data results from the contribution of the heart beat 
to physiological tremor (Elble and Koller 1977). This additional 
influence on the ACC apart from the EMG is not captured 
by the model. Thus, the coherency of the measured data is reduced more
than expected from the model that only considers the effect of
observational noise. 

From a comparison of Fig.~\ref{flat_emg_fig}d and \ref{flat_emg_fit}d
it can be concluded that the small oscillations of the
cross-correlation for negative lags give no evidence for a reflex feedback.
These oscillations appear because the
cross-correlation estimates are not uncorrelated
as discussed in Section~3.3. 

Assuming the validity of the AR[2] model to describe the physiological 
tremor, one can compare the peak frequency estimated from the power spectrum
 (\ref{peakfreq}) 
with that estimated from the phase spectrum according to
({\ref{a1},{\ref{a2},\ref{ar2phase}). 
Taking the errors of the estimates into account both values are
consistent. 

We received similar results in 70\% of the investigated
series. In the other 30\% of the cases an interpretation of the phase spectrum
was not possible because of the poor coherency and, therefore, large errors
in the phase spectrum. 
As discussed in Section~3 this might be simply the result of a smaller
signal-to-noise ratio in the EMG and/or ACC in these cases due to a very
low tremor amplitude. An estimation of the power in the ACC and 
EMG spectra supported this hypothesis.

\section {Summary} \label {summary}
We investigated the relation between muscle activity and  physiological
tremor in the case of unsynchronized EMG activity by cross-spectral 
analysis with special respect to the phase spectrum and the effects of
observational noise.
Such an analysis is not a straightforward task since one can not 
a priori expect a one to one relationship between one muscle and a specific
mechanical measure like force, movement or acceleration due to
the redundancy of the muscle system. Furthermore, the analysis
is handicapped by the small tremor amplitude.

We found that this type of physiological 
tremor can be regarded as an AR process driven by the uncorrelated EMG
activity without involving any reflex mechanisms. 
We showed that the phase spectrum between EMG and ACC
can not be interpreted at a single frequency in terms of a delay.
The phase spectrum depends on the
mechanical properties of the hand, i.e.~the driven part of the
system, but not on the characteristics of the driving force.
The behavior of the coherency spectrum can be explained as an effect
of the frequency dependent signal-to-noise ratio. In addition, 
for low frequencies, the effect of the heart beat on the tremor
further reduces the coherency between the EMG and the ACC.

Autoregressive processes of order 2 are derived from stochastic
differential equations where the noise represents a force 
in a Newtonian sense, i.e.~causing an acceleration.
The conformity of the theoretical phase spectrum assuming such a process
with the empirical data shows that in the case of this
small amplitude hand tremor the measured EMG represents 
a Newtonian force by which the muscle acts on the hand.

\section*{References}
\begin{itemize}
\item [] Bloomfield P (1976) Fourier analysis of time series: An
         introduction. John Wiley. New York

\item [] Bartlett MS (1978) Stochastic Processes. Cambridge University
         Press, Cambridge

\item [] Brockwell PJ, Davis RA (1987) Time Series: Theory and
         Methods. Springer, Berlin

\item [] Deuschl G, Blumberg H, L\"ucking CH (1991) Tremor in reflex
         sympathetic dystrophy. Arch Neurol 48:1247-1252

\item [] Elble RJ, Koller WC (1977) Mechanistic components of normal
        hand tremor. Electroencephal clin Neurophysiol 44:72-82

\item [] Elble RJ, Koller WC (1990) Tremor. The Johns Hopkins
         University Press, Baltimore
 
\item [] Elble RJ, Randall JE (1976) Motor-unit activity responsible
         for 8- to 12-Hz component of human physiological finger tremor. 
         J Neurophys 39:370-383

\item [] Fox JR, Randall JE (1970) Relationship between forearm tremor
         and the biceps electromyogram. J Appl Neurophys 29:103-108

\item [] Gantert C, Honerkamp J, Timmer J (1992) Analyzing the
         dynamics of hand tremor time series. Biol Cybern 66:479-484

\item [] Hannan EJ, Thomson PJ (1971) The estimation of coherency and group
         delay. Biometrika 58:469-481

\item [] Honerkamp J (1993) Stochastic dynamical systems. VCH, New York

\item [] Iaizzo PA, Pozos RS (1992) Analysis of multiple EMG and 
         acceleration signals of various record length as a means to study
         pathological and physiological oscillations. Electromyogr clin
         Neurophysiol 32:359-367

\item [] Journee HL (1983) Demodulation of amplitude modulated noise:
         A mathematical evaluation of a demodulator for pathological tremor EMG's.
         IEEE Trans Biomed Eng 30:304-308
         
\item [] K\"onig M, Timmer J (1997) Analyzing X-ray variability by linear
         state space models, Astronomy Astophys SS 124, 589-596

\item [] Lenz FA, Zasker RR, Kwan HC, Schnider S, Kwong R, Murayama Y, 
         Dostrovsky JO, Murphy JT (1988) Single unit analysis of the human 
         ventral thalamic nuclear group: correlation of thalamic ''tremor 
         cells'' with the 3-6 Hz component of parkinsonian tremor.
         J Neuroscience 8:754-764

\item []  Miao T. and Sakamoto K. (1995) Effects of weight load on
          physiological tremor: The AR representation. Applied Human
          Science 14:7-13

\item [] Pashda SM, Stein RB (1973) The bases of tremor during a
         maintained posture. In: Control of Posture and Locomotion, Ed.: RB
         Stein, KG Pearson, RS Smith, JB Redford. New York. Plenum. pp 415-419

\item [] Press HP, Teukolsky SA, Vetterling WT, Flannery BP. 1992.
         Numerical Recipes, Cambridge University Press, Cambridge

\item [] Priestley MB (1989) Spectral Analysis and Time Series.
         Academic Press

\item [] Randall JE (1973) A Stochastic Time Series Model for Hand
         Tremor. J Appl Physiol 34:390-395

\item [] Rietz RR, Stiles RN (1974) A viscoelastic mass mechanism as a basis 
         for normal postural tremor. J Appl Physiol 37:852-860 

\item [] Stiles NS, Randall JE (1967) Mechanical factors in human tremor
         frequency. J Appl Physiol 23:324-330

\item [] Stiles NS (1980) Mechanical and Neural feedback factors in
         postural hand tremor of normal subjects. J Neurophys 44:40-59

\item [] Stiles NS (1983) Lightly damped hand oscillations:
         Acceleration-related feedback and system damping. J Neurophys
         50:327-343 

\item [] Timmer J, Gantert C, Deuschl G, Honerkamp J (1993)
         Characteristics of hand tremor time series. Biol Cybern 70:75-80

\item [] Timmer J, Lauk M, Deuschl G (1996) Quantitative analysis
         of tremor time series. Electroencephal clin Neurophys 101:461-468

\item [] Timmer J, Lauk M,  Pfleger W, Deuschl G (1998)
         Cross-spectral analysis of physiological tremor and muscle 
         activity. II. Application to synchronized EMG, submitted to 
         Biol Cybern

\item [] Timmer J, Lauk M, L\"ucking CH (1997)
         Confidence regions for spectral peak frequencies.
         Biometr J 39:849-861

\end {itemize}
\end {document}